\documentclass[aps,prl,twocolumn,superscriptaddress,reprint,longbibliography]{revtex4-2}

\usepackage{graphicx}
\usepackage{dcolumn}

\usepackage{bm}
\usepackage{siunitx}
\usepackage{amssymb}
\usepackage[hidelinks]{hyperref}

\usepackage{xcolor}

\usepackage{amsmath}

\graphicspath{{Figures/}}

\newcommand{\rperm}{\varepsilon_\mathrm{r}}

\begin{document}

\preprint{APS/123-QED}

\title{Overscreening and Underscreening in Solid-Electrolyte Grain Boundary Space-Charge Layers}

\author{Jacob~M.~Dean}
\affiliation{Department of Chemistry, University of Bath, Claverton Down BA2 7AY, United Kingdom}
\affiliation{The Faraday Institution, Quad One, Harwell Science and Innovation Campus, Didcot OX11 0RA, United Kingdom}
\author{Samuel~W.~Coles}
 \email{swc57@bath.ac.uk}
\affiliation{Department of Chemistry, University of Bath, Claverton Down BA2 7AY, United Kingdom}
\affiliation{The Faraday Institution, Quad One, Harwell Science and Innovation Campus, Didcot OX11 0RA, United Kingdom}
\author{William~R.~Saunders}
\affiliation{Department of Physics, University of Bath, Claverton Down BA2 7AY, United Kingdom}
\author{Andrew~R.~McCluskey}
\affiliation{Data Management and Software Centre, European Spallation Source ERIC, Ole Maal\o es vej 3, 2200 K\o benhavn, Denmark}
\affiliation{Department of Chemistry, University of Bath, Claverton Down BA2 7AY, United Kingdom}
\author{Matthew~J.~Wolf}
\affiliation{Department of Physics, University of Bath, Claverton Down BA2 7AY, United Kingdom}
\author{Alison~B.~Walker}
\affiliation{Department of Physics, University of Bath, Claverton Down BA2 7AY, United Kingdom}
\author{Benjamin~J.~Morgan}
 \email{b.j.morgan@bath.ac.uk}
\affiliation{Department of Chemistry, University of Bath, Claverton Down BA2 7AY, United Kingdom}
\affiliation{The Faraday Institution, Quad One, Harwell Science and Innovation Campus, Didcot OX11 0RA, United Kingdom}

\date{\today}

\begin{abstract}
Polycrystalline solids can exhibit material properties that differ significantly from those of equivalent single-crystal samples, in part, because of a spontaneous redistribution of mobile point defects into so-called space-charge regions adjacent to grain boundaries.
The general analytical form of these space-charge regions is known only in the dilute limit, where defect--defect correlations can be neglected.
Using kinetic Monte Carlo simulations of a three-dimensional Coulomb lattice gas, we show that grain-boundary space-charge regions in non-dilute solid electrolytes exhibit overscreening---damped oscillatory space-charge profiles---and underscreening---decay lengths that are longer than the corresponding Debye length and that increase with increasing defect--defect interaction strength.
Overscreening and underscreening are known phenomena in concentrated liquid electrolytes, and the observation of functionally analogous behaviour in solid electrolyte space-charge regions suggests that the same underlying physics drives behaviour in both classes of systems.
We therefore expect theoretical approaches developed to study non-dilute liquid electrolytes to be equally applicable to future studies of solid electrolytes.
\end{abstract}

\keywords{underscreening, overscreening, grain boundary, space-charge, solid electrolyte}

\maketitle


Predicting the equilibrium distribution of charged ions within crystalline solids at, or near, structural discontinuities such as grain boundaries or heterointerfaces is a long-standing problem in solid-state physics \cite{Frenkel_KineticTheoryOfLiquids1946, EshelbyEtAl_PhilMag1958, KliewerAndKoehler_PhysRev1965, PoeppelAndBlakely_SurfSci1969, ChengEtAl_Joule2020,SwiftEtAl_NatureComputSci2021}.
Within these near-interface regions, the concentration of individual ionic species can deviate significantly from their formal bulk values, giving rise to so-called ``space-charge'' regions.
The spatial profiles of space-charge regions is particularly significant in solid electrolytes \cite{Maier_ProgSolStatChem1995,XuEtAl_NatureMater2020}, such as those used in solid-oxide fuel cells and all--solid-state lithium ion batteries \cite{AdepalliEtAl_AdvFuncMater2017,deKlerkAndWagemaker_ACSApplEnergyMater2018,ChengEtAl_Joule2020,SwiftEtAl_NatureComputSci2021}.
In these cases a local decrease in the concentration of the charge-carrying mobile ionic species within a space-charge region is expected to contribute to interfacial resistance and decreased device performance \cite{Maier_ProgSolStatChem1995, HelgeeEtAl_FuelCells2012, UthayakumarEtAl_JPhysChemC2020}.

The classic treatment of space-charge formation in solid electrolytes considers the distribution of mobile ions in terms of charge-carrying point defects---typically interstitials or vacancies---that behave as an ideal gas interacting only through mean-field electrostatics \cite{Maier_ProgSolStatChem1995}.
For this simple model, the equilibrium defect distribution perpendicular to the grain-boundary interface can be found by solving the one-dimensional Poisson--Boltzmann equation.
For low space-charge potentials \footnote{The low--space-charge potential limit is formally given by $\Phi_\mathrm{sc} q\ll k_\mathrm{B}T$, where $\Phi_\mathrm{sc}$ is the ``space-charge potential'' at the interface, $q$ is the charge of the mobile defect species, $k_\mathrm{B}$ is the Boltzmann constant, and $T$ is the temperature.}, the approximate linearised Poisson--Boltzmann equation can be used, which has a general analytical solution;
the resulting space-charge profiles decay exponentially towards the asymptotic bulk defect concentration with a characteristic decay length equal to the Debye length, $\lambda_\mathrm{D}$;
\begin{equation}
\lambda_\mathrm{D} = \sqrt{\frac{\varepsilon_{0}\rperm k_\mathrm{B}T}{\sum_{i} (z_{i}e)^{2}c_{i}}}, 
\label{eqn:debye_length}
\end{equation}
where $\varepsilon_{0}$ is the vacuum permittivity, $\rperm$ is the relative permittivity of the solid, $k_\mathrm{B}$ is the Boltzmann constant, $T$ is the temperature, $z_{i}$ are the charged species' valences, $e$ is the elementary charge, and $c_{i}$ are the bulk charged defect concentrations.
For high space-charge potentials the full non-linear Poisson--Boltzmann equation must be solved, either numerically or by making further approximations to obtain an approximate analytical result \cite{TongEtAl_JAmCeramSoc2020}. Space-charge profiles for high space-charge potentials again decay monotonically, exhibiting superexponential decay close to the grain-boundary, converging to exponential decay with decay length $\lambda_\mathrm{D}$ at larger distances.

This ideal-gas plus mean-field--electrostatics model of space-charge behaviour is valid only in the dilute limit of low defect density or high relative permittivity, i.e.~weak electrostatic coupling.
To address this limitation, a number of more complex one-dimensional models have been proposed that aim to describe space-charge behaviour under non-dilute conditions \cite{Mebane_CompMaterSci2015,PatsahanEtAl_SolStatIonics2019,TongEtAl_JAmCeramSoc2020,MebaneAndDeSouza_EnergyEnivronSci2015,deKlerkAndWagemaker_ACSApplEnergyMater2018,BondevikEtAl_SolStatIonics2020}.
These models extend the simple model described above by including additional non-ideal local or non-local defect chemical potential terms.
The equilibrium space-charge profile is then computed by minimising the global free energy as a function of the one-dimensional defect concentration.  
These extended models can give complex non-exponential or even non-monotonic space-charge profiles, although the exact functional form of the resulting profile depends on the choice of non-ideal chemical potential terms and their parameterisation. 

As is the case for the classic space-charge model, these non-dilute space-charge models solve a one-dimensional problem defined in terms of the mobile-defect density perpendicular to the interfacial plane.
Each point in this one-dimensional density can be considered a planar average from some corresponding implicit three-dimensional defect distribution; the full three-dimensional distribution of mobile defects within the solid-electrolyte host, however, is never explicitly calculated.
As a consequence, defect--defect correlations, which become increasingly important with increasing defect concentration or decreasing relative permittivity, are not explicitly computed. Instead, the effect of these correlations is approximated through the choice of non-ideal defect chemical potential terms used in each model.

Here, we follow a different approach to modelling solid-electrolyte space-charge profiles.
Instead of solving an effective one-dimensional problem, we consider an explicit three-dimensional Hamiltonian for point charges in a system with a single grain boundary. 
We then sample the configuration space of this model, using kinetic Monte Carlo, and construct space-charge profiles as time-averages over specific simulation trajectories.
Because this approach treats defect--defect interactions explicitly in three-dimensional space, any defect--defect correlations emerge directly from the simulation trajectories, avoiding the need to describe these correlations through analytical free-energy terms.

We find that in non-dilute solid electrolytes (high defect concentrations) or for strong defect--defect interactions (low relative permittivities) grain-boundary--adjacent space-charge profiles have a qualitatively different functional form to the monotonic decay predicted by classic space-charge theory.
We observe damped oscillatory space-charge profiles that can be well described by a simple analytical function.
We also observe space-charge decay lengths that are significantly larger than the corresponding Debye length, and that increase with increasing defect concentration or defect--defect interaction strength (decreasing relative permittivity), giving the opposite behaviour to that predicted by dilute-limit models.
The deviation of space-charge decay length from the Debye length is shown to follow a universal scaling law as a function of electrostatic coupling strength.
We note that analogous phenomena have been previously reported for concentrated liquid electrolytes, which suggests that the same underlying physics drives emergent behaviour in solid electrolytes as in their liquid counterparts.
Finally, we comment on the implications of these results for a possible unified theoretical framework for understanding non-dilute electrolyte behaviour in both liquids and solids.

\begin{figure}[tb]
  \centering
  \resizebox{6.3cm}{!}{\includegraphics*{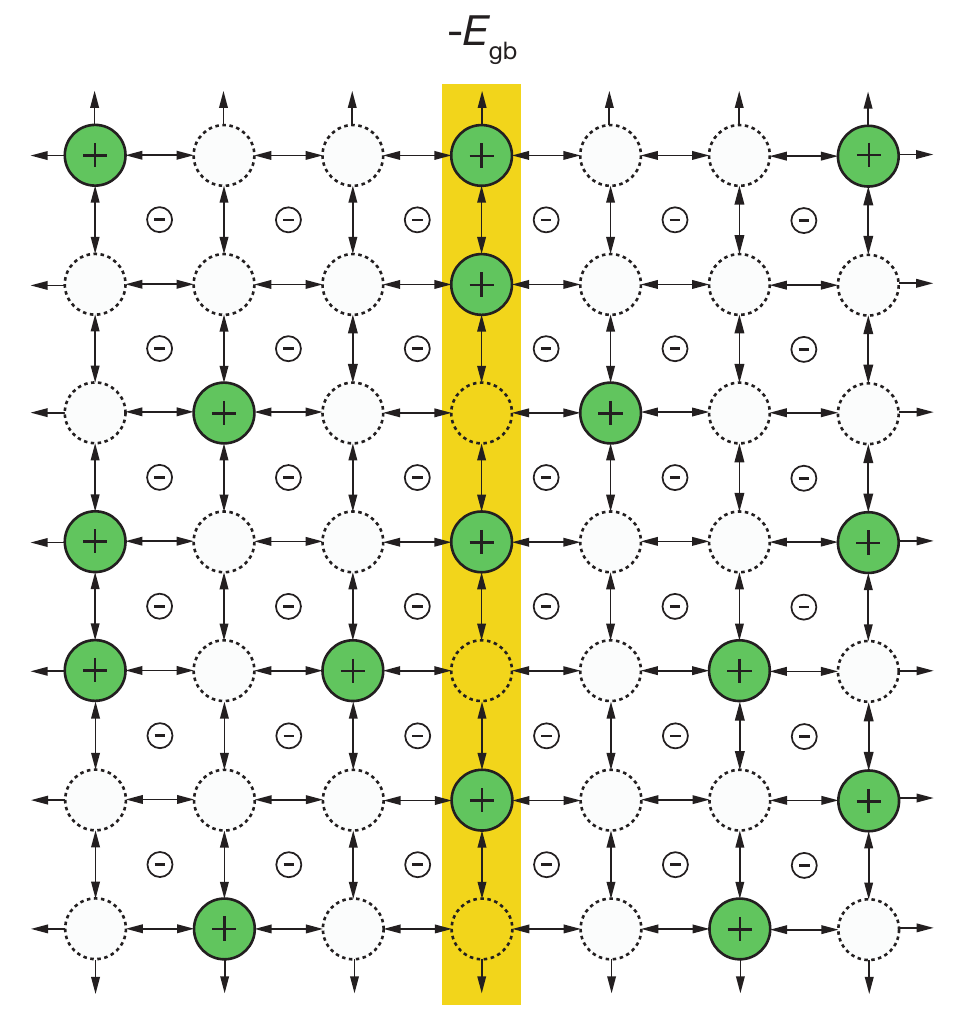}} %
    \caption{\label{fig:schematic}Two-dimensional schematic of the kinetic Monte Carlo model used in this work. Lattice sites (circles) are either vacant (dashed circles) or occupied by positively charged ``defects'' (solid green circles). Arrows indicate allowed site--site moves. Interstitial positions are assigned partial negative charges. All lattice sites in the central plane (yellow) are assigned an on-site occupation energy of $-E_\mathrm{gb}$.}
\end{figure}

\emph{Methods:} Our computational model consists of a 3D-periodic simple-cubic lattice of $m\times m\times m$ sites, populated by a fixed number, $N$, of mobile point defects with $+1$ charge (Fig.~\ref{fig:schematic}).
To ensure our model has net zero charge, we include a uniform array of $\delta^-$ point charges located at the cube-centre--interstitial positions, with $\delta^-=-N/m^3$. The mobile defects and their counter-charges interact through point-charge electrostatics, which we scale by a relative permittivity $\rperm$.
This Coulomb lattice-gas model has historically been well studied as a bulk system \cite{WalkerAndGillan_JPhysCSolStatPhys1983}, particularly in the context of the classical one-component plasma \cite{BrushEtAl_JChemPhys1966,BausAndHansen_PhysRep1980}.
To model the segregation of mobile defects to a grain boundary, we assign one  plane of sites an on-site occupation energy of $-E_\mathrm{gb}$; this term represents the difference in standard chemical potential or ``segregation energy'' for defects in the grain boundary core relative to the bulk. 
All sites outside the grain-boundary core have on-site energies of zero.

\begin{figure*}[th]
  \centering
  \resizebox{17.8cm}{!}{\includegraphics*{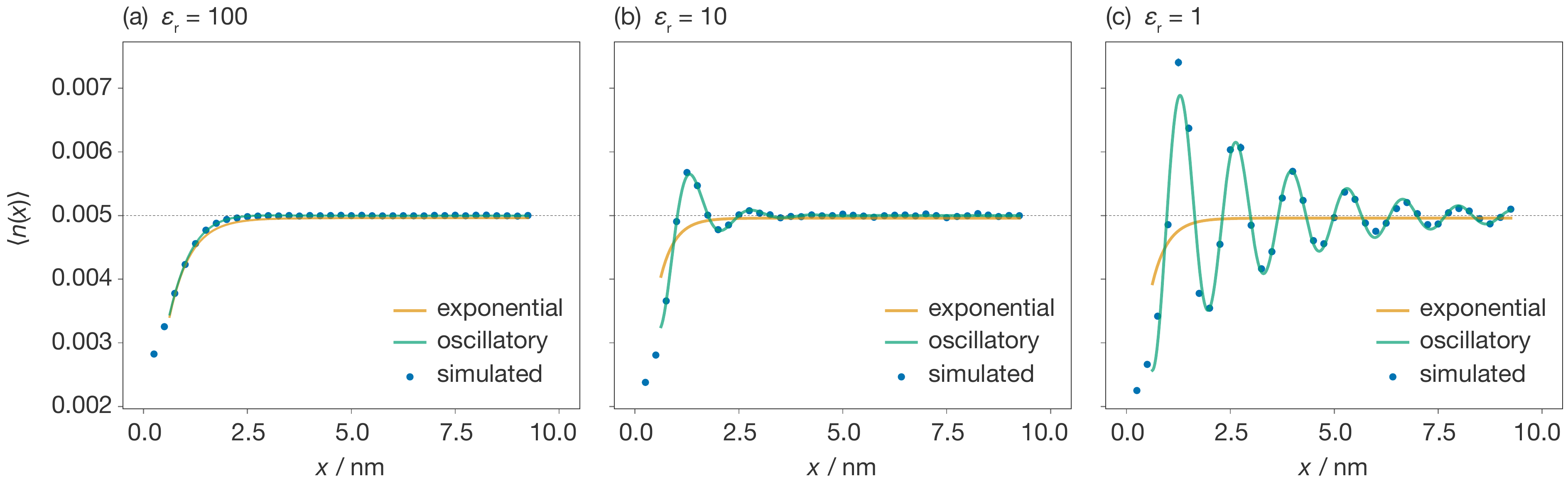}} %
    \caption{\label{fig:overscreening}One-dimensional time-averaged mobile defect distributions for $n_{\infty} = 0.005$ and relative permittivities (a) $\rperm =100$, (b) $\rperm=10$, and (c) $\rperm=1$. On each plot, $x=0$ corresponds to the grain-boundary plane.
    For each set of simulation data we also plot the maximum likelihood exponential and oscillatory models (Eqns.~\ref{eqn:exponential} and \ref{eqn:oscillatory_model} respectively). Error bars show a 95\% confidence interval around each point (errors are smaller than the symbols). 
    Plots showing these data on a semi-logarithmic scale are provided in the  Supporting Information. \emph{Source:} Raw simulation data and scripts to generate this figure are available under CC BY 4.0 / MIT licences as Ref.~\onlinecite{Dean_dataset}.}
\end{figure*}

Our kinetic Monte Carlo simulations were performed using the \texttt{kmc\_fmm} code, which implements the accurate and optimally scaling electrostatic solver described in Ref.~\onlinecite{SaundersEtAl_JCompPhys2020} using the \texttt{ppmd} framework \cite{SaundersEtAl_CompPhysCommun2018}.
All simulations were performed using a $75 \times 75 \times 75$ simple-cubic lattice of sites, with three-dimensional periodic boundary conditions.
The nearest-neighbour mobile-site spacing is \SI{2.5}{\angstrom}, which approximates the typical lattice spacing of a solid electrolyte.
For all simulations, we set the on-site occupation energy for the grain-boundary plane as $E_\mathrm{gb}=\SI{-0.05}{eV}$.
Simulations were run at $\SI{300}{K}$.
Each individual simulation was initialised with a random distribution of mobile charges and run for a total of 1,500,000 steps. 
The initial 500,000 steps were discarded to allow for equilibration, and space-charge profiles were computed as a time-average over the subsequent 1,000,000 steps. 
For each combination of relative permittivity, $\rperm$, and bulk concentration, $n_\infty$, we performed between 30 and 100 simulations, to obtain statistically converged space-charge profiles.
To characterise and quantify the simulated space-charge profiles we performed maximum likelihood sampling, parameter posterior sampling, and Bayesian model selection for competing functional forms using the \texttt{uravu} Python package \cite{McCluskey2020}. The posterior distributions and Bayesian evidence for specific functional models were found by nested sampling \cite{Skilling2004}, implemented in \texttt{uravu} using the \texttt{dynesty} package \cite{Speagle2020}. Further details are given in the Supporting Information.

\emph{Results:} We first consider simulated space-charge profiles obtained for $n_{\infty} = 0.005$, where $n_\infty$ is the bulk per-site number density of mobile defects, and $\rperm=\numlist[list-final-separator={\text{ and }}]{100;10;1}$ (Fig.~\ref{fig:overscreening}), which spans the range of relative permittivities for typical solid electrolytes \cite{deKlerkAndWagemaker_ACSApplEnergyMater2018}.
At higher values of $\rperm$ the electrostatic interaction between the mobile defects is more effectively screened, and each model will more closely approximate the non-interacting (dilute) ideal lattice-gas limit. 
Conversely, as $\rperm$ is decreased the electrostatic interactions between mobile defects become stronger and non-ideal defect--defect correlations are expected to become more significant.

For the highest permittivity considered here ($\rperm=100$) we obtain a monotonically decaying space-charge profile (Fig.~\ref{fig:overscreening}(a)) that is qualitatively consistent with the dilute-limit behaviour predicted by the classic Poisson--Boltzmann model.
More precisely, the simulated space-charge profile is well described by an exponentially decaying function
\begin{equation}
n(x) = n_\infty + A\exp(-\alpha x),
\label{eqn:exponential}
\end{equation}
where $n(x)$ is the planar-average defect number density, $n_\infty$ is the bulk defect number density, and $\alpha=1/\lambda_\mathrm{sys}$, with $\lambda_\mathrm{sys}$ the observed characteristic screening length.
At lower permittivities the simulated space-charge profiles are no longer monotonic (Fig.~\ref{fig:overscreening}(b)), and are therefore not even qualitatively well-described by Eqn.~\ref{eqn:exponential}. 
We instead observe oscillatory behaviour that decays into the bulk.
Further decreases in $\rperm$ cause both the magnitude of these oscillations and the distance over which they decay to increase (Fig.~\ref{fig:overscreening}(c)).

The appearance of oscillations at higher defect--defect interaction strengths (decreasing $\rperm$) mirrors the behaviour of non-dilute liquid electrolytes and ionic liquids, which exhibit similar oscillatory charge profiles at electrified interfaces, where this phenomenon is termed ``overscreening''  \cite{BazantEtAl_PhysRevLett2011}.
Oscillatory local order has also been observed in previous bulk Coulomb lattice-gas and one-component plasma simulations \cite{BrushEtAl_JChemPhys1966,BausAndHansen_PhysRep1980,PetersenAndDietrich_PhilMagB1992}.
By analogy to the analytically-known behaviour of liquid systems \cite{LeeEtAl_PhysRevLett2017,LeeEtAl_FaradayDiscuss2017}, we propose a damped oscillatory ansatz for the space-charge profiles in this low-permittivity regime:
\begin{equation}
n(x) = n_\infty + A\exp(-\alpha x)\cos(\xi x + \theta).
\label{eqn:oscillatory_model}
\end{equation} 
As shown in Figs.~\ref{fig:overscreening}(b) and \ref{fig:overscreening}(c), this oscillatory functional form gives a significantly better description of these space-charge profiles at low permittivities than the dilute-limit monotonic (purely exponential) model \footnote{The possibility of solid electrolytes exhibiting oscillatory charge-profiles was previously highlighted by Gray-Weale \cite{Gray-Weale_FaradayDiscuss2007}, although this earlier analysis did not resolve the decay behaviour of these oscillations.}.

The simulated space-charge profiles presented in Fig.~\ref{fig:overscreening} also show \emph{increasing} space-charge widths---i.e.\ increased screening lengths---with \emph{decreasing} relative permittivities.
This is the opposite behaviour to that predicted by the classic space-charge treatment, wherein the Debye length decreases with decreasing $\rperm$ (Eqn.~\ref{eqn:debye_length}).
The observation of a screening length that is larger than the classical Debye length again mirrors the behaviour of concentrated liquid electrolytes, where this phenomenon is termed ``underscreening'' \cite{LeeEtAl_PhysRevLett2017,LeeEtAl_FaradayDiscuss2017,Limmer_PhysRevLett2015}.

In ionic liquids and concentrated liquid electrolytes, underscreening has been found to follow scaling laws of the form
\begin{equation}
\left(\frac{\lambda_\mathrm{sys}}{\lambda_\mathrm{D}}\right) \propto \left(\frac{d}{\lambda_\mathrm{D}}\right)^{\nu},
\label{eqn:universal_scaling}
\end{equation}
where $\lambda_\mathrm{sys}$ is the observed screening length of the system, $d$ is the diameter of the charged species, and $\nu$ is a fixed scaling factor \cite{Attard_PhysRevE1993,SmithEtAl_JPhysChemLett2016,AdarEtAl_PhysRevE2019,ColesEtAl_JPhysChemB2020}.
For our Coulomb lattice-gas model there is no direct analogue for $d$.
We instead analyse our simulation data by plotting $
\ln\left(\lambda_\mathrm{sys}/\lambda_\mathrm{D}\right)$ versus $\ln(\Gamma)$, where $\Gamma$ is a dimensionless parameter that describes the strength of electrostatic coupling in a given system. 
$\Gamma=\lambda_\mathrm{B}/a$, where $\lambda_\mathrm{B}=e^2/(4 \pi \varepsilon_0 \varepsilon_\mathrm{r} k_\mathrm{B} T)$ is the Bjerrum length, and $a=\left[3/(4\pi n_\infty)\right]^{1/3}$ is a Wigner--Seitz defect-sphere radius \cite{Carley_PhysRev1963,BrushEtAl_JChemPhys1966,BausAndHansen_PhysRep1980}.
Further discussion of the choice of an appropriate scaling parameter is provided in the SI.

\begin{figure}[t]
  \centering
  \resizebox{7.0cm}{!}{\includegraphics*{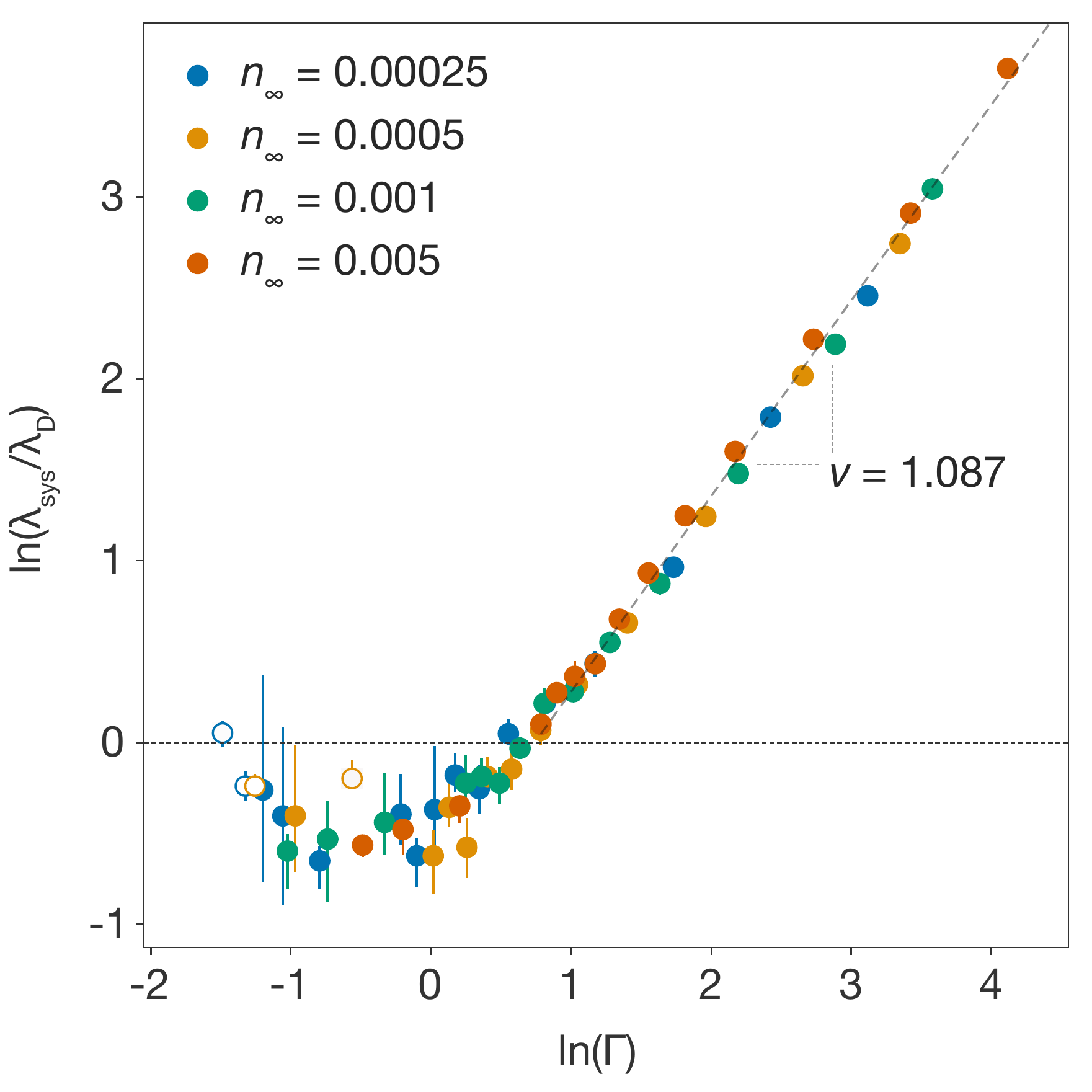}} %
    \caption{\label{fig:underscreening} Log--log plot of scaled simulated decay length $\lambda_\mathrm{sys}/\lambda_\mathrm{D}=1/(\alpha\lambda_\mathrm{D})$ versus $\Gamma=\lambda_\mathrm{B}/a$ for simulations performed at four different defect concentrations. The diagonal dashed line shows the maximum likelihood estimate for Eqn.~\ref{eqn:universal_scaling} in the underscreening regime. Open and closed circles correspond to space-charge profiles that best support the pure exponential decay model (Eqn.~\ref{eqn:exponential}) or the oscillatory decay model (Eqn.~\ref{eqn:oscillatory_model}) respectively, as determined through Bayesian model selection (see SI for details). Error bars show 95\% confidence intervals for $\lambda_\mathrm{sys}/\lambda_\mathrm{D}$. \emph{Source:} Raw simulation data and scripts to generate this figure are available under CC BY 4.0 / MIT licences as Ref.~\cite{Dean_dataset}.}
\end{figure}

Fig.~\ref{fig:underscreening} shows $
\ln\left(\lambda_\mathrm{sys}/\lambda_\mathrm{D}\right)$ versus $\ln(\Gamma)$ plotted for simulations at four bulk concentrations: $n_\infty = 0.00025$, $0.0005$, $0.001$, and $0.005$. 
For each simulation, we use Bayesian model selection to select which model---purely exponential or damped oscillatory---is best supported by the simulated space-charge--profile data, and assign $\lambda_\mathrm{sys} = 1/\alpha$. 
For $\ln(\Gamma)>0.4$, we find analogous scaling behaviour to that reported for liquid systems \cite{LeeEtAl_PhysRevLett2017,ColesEtAl_JPhysChemB2020, RotenbergEtAl_JPhysCondensMatter2018,AdarEtAl_PhysRevE2019}, i.e.~$\lambda_\mathrm{sys}/\lambda_\mathrm{D} \propto \Gamma^\nu$, with $\nu=1.087\pm0.005$ (95\% CI). 
For a given bulk concentration of mobile defects, higher values of $\Gamma$ correspond to stronger defect--defect interactions, i.e.\ decreased $\rperm$.
In this regime, as defect--defect interactions increase in magnitude the observed screening length increasingly positively deviates from the Debye length. 
For $\ln(\Gamma)<0.4$ we observe a region where the observed screening length is \emph{smaller} than the Debye length.
Again, this behaviour has been reported for liquid systems in studies that focus on the application of different closure relationships to the Ornstein--Zernike equation \cite{OZ, RotenbergEtAl_JPhysCondensMatter2018,Attard_PhysRevE1993,CoupetteEtAl_PhysRevLett2018,AdarEtAl_PhysRevE2019,Kirkwood_ChemRev1936,ennis_dressed_1995, Kjellander_SoftMatt2019,kjellander_focus_2018,cats2020primitive}.
At the lowest values of $\Gamma$ ($\approx0.25$, obtained for $n_\infty=0.00025$ and $\rperm=100$) we recover the low-potential dilute-limit result that the simulated screening length is equal to the Debye length; $\lambda_\mathrm{sys}=\lambda_\mathrm{D}$.
We note that the same scaling behaviour is observed for all four simulated concentrations, suggesting a universal scaling law operates in solid electrolytes that contain a single mobile defect species.

\emph{Summary and Discussion:} We have performed kinetic Monte Carlo simulations of populations of mobile point defects on a three-dimensional lattice, interacting via point-charge electrostatics, as a simple model for space-charge formation at solid-electrolyte grain-boundaries, for a range of bulk defect concentrations and relative permittivities. 
For dilute, weakly coupled systems (low particle concentration and high $\rperm$) we recover the behaviour predicted by classic treatments of space-charge behaviour, i.e.~the linearised Poisson--Boltzmann model---the space-charge profile decays exponentially to the asymptotic bulk value with a decay length equal to the Debye length $\lambda_\mathrm{D}$.
In non-dilute, strongly coupled systems (high particle concentration or low $\rperm$) we observe damped oscillatory space-charge profiles---overscreening---and space-charge decay lengths that are larger than the corresponding Debye length---underscreening.

Overscreening and underscreening are known phenomena for non-dilute liquid systems of charged particles (concentrated liquid electrolytes and ionic liquids) \cite{BazantEtAl_PhysRevLett2011, FedorovAndKornyshev_ChemRev2014,Kornyshev_JPhysChemB2007,PerkinEtAl_PhysChemChemPhys2010,SmithEtAl_JPhysChemLett2016,LeeEtAl_FaradayDiscuss2017,LeeEtAl_PhysRevLett2017,GebbieEtAl_ProcNatAcadSci2013,GebbieEtAl_ProcNatAcadSci2015,EvansAndSluckin_MolPhys1980,LeotedeCarvalhoAndEvans_MolPhys1994,EvansEtAl_MolPhys1993}.
Our simulations predict not only that overscreening and underscreening also occur in non-dilute solid electrolytes, but that these phenomena have the same functional behaviour in solids as in analogous liquid systems.
Space-charge profiles in systems that exhibit overscreening are well described by a damped oscillatory function and the degree of underscreening, given by the ratio of the observed space-charge decay length to the classic Debye length, follows an apparently universal scaling law above some critical effective defect--defect interaction strength.
We also predict an intermediate regime, where the simulated space-charge decay length is \emph{smaller} than the classic Debye length.
Again, this mirrors previous results obtained for the decay of radial distribution functions in non-dilute liquid systems \cite{RotenbergEtAl_JPhysCondensMatter2018, AdarEtAl_PhysRevE2019, Attard_PhysRevE1993,cats2020primitive}.
Our results also suggest an approximate empirical threshold for when the dilute-limit classic space-charge treatment may be reliably used to quantitatively model space-charge behaviour in solid electrolytes of $\Gamma=\lambda_\mathrm{B}/a\lessapprox0.25$.

The strong correspondence between our results for space-charge formation at a grain boundary in a model solid electrolyte and those previously reported for liquid electrolytes at an electrified interface suggests that the same underlying physics is responsible for these analogous emergent phenomena in both liquid and solid electrolytes.
This suggests that many of the theoretical methods that have been developed previously by researchers seeking to better understand the behaviour of liquid electrolytes, or analogous models such as the one-component plasma \cite{BrushEtAl_JChemPhys1966,BausAndHansen_PhysRep1980}, may be equally applicable to developing an improved understanding of the emergent behaviour of ensembles of mobile charged defects in solid electrolytes.
\\

\emph{Acknowledgements}
J.~M.~D.\ acknowledges support from the EPSRC (grant number 2119790).  J.~M.~D.\ and B.~J.~M.\ acknowledge the support of the Faraday Institution through the Multi-Scale Modelling project (grant no. FIRG003).
S.~W.~C.\ and B.~J.~M.\ acknowledge the support of the Faraday Institution through the CATMAT project (grant number FIRG016).
B.~J.~M.\ acknowledges support from the Royal Society (UF130329 \& URF\textbackslash R\textbackslash 191006).
W.~R.~S., M.~J.~W., and A.~B.~W.\ acknowledge funding from the European Union’s Horizon 2020 programme, via the Energy Oriented Centre of Excellence (EoCoE-II), under grant agreement number 676629. The authors thank Dr. Ian R.\ Thompson for useful discussions.
We are grateful to the UK Materials and Molecular Modelling Hub for computational resources, which is partially funded by EPSRC (EP/P020194/1). This research made use of the Balena High Performance Computing (HPC) Service at the University of Bath.
\\

\emph{Supporting Information}
  Discussion of Fig.~\ref{fig:overscreening} when replotted on a semi-logarithmic scale; discussion of the choice of non-dimensional coupling parameter used in Fig.~\ref{fig:underscreening}; further details of the Bayesian analysis of the simulated space-charge profiles and subsequent parameter estimation.
  \\

\emph{Data availability:}
A dataset containing the full set of time-averaged simulation data that support the findings presented here, and the analysis code used to generate Fig.~\ref{fig:overscreening} and Fig.~\ref{fig:underscreening} is available as Ref.~\onlinecite{Dean_dataset} under the CC BY 4.0 and MIT licences.

\bibliography{Jacob}

\end{document}


\setchemformula{
  radical-radius = .3ex , 
  charge-hshift  = 0pt    
}
\title{Supplementary Information for ``Overscreening and Underscreening in Solid-Electrolyte Grain Boundary Space-Charge Layers"}

\author{Jacob~M.~Dean}
\affiliation{Department of Chemistry, University of Bath, Claverton Down BA2 7AY, United Kingdom}
\affiliation{The Faraday Institution, Quad One, Harwell Science and Innovation Campus, Didcot, United Kingdom}
\author{Samuel~W.~Coles}
 \email{swc57@bath.ac.uk}
\affiliation{Department of Chemistry, University of Bath, Claverton Down BA2 7AY, United Kingdom}
\affiliation{The Faraday Institution, Quad One, Harwell Science and Innovation Campus, Didcot, United Kingdom}
\author{William~R.~Saunders}
\affiliation{Department of Physics, University of Bath, Claverton Down BA2 7AY, United Kingdom}
\author{Andrew~R.~McCluskey}
\affiliation{Data Management and Software Centre, European Spallation Source ERIC, Ole Maal\o es vej 3, 2200 K\o benhavn, Denmark}
\affiliation{Department of Chemistry, University of Bath, Claverton Down BA2 7AY, United Kingdom}
\author{Matthew~J.~Wolf}
\affiliation{Department of Physics, University of Bath, Claverton Down BA2 7AY, United Kingdom}
\author{Alison~B.~Walker}
\affiliation{Department of Physics, University of Bath, Claverton Down BA2 7AY, United Kingdom}
\author{Benjamin~J.~Morgan}
 \email{b.j.morgan@bath.ac.uk}
\affiliation{Department of Chemistry, University of Bath, Claverton Down BA2 7AY, United Kingdom}
\affiliation{The Faraday Institution, Quad One, Harwell Science and Innovation Campus, Didcot, United Kingdom}

\date{\today}

\begin{abstract}
This document presents supplementary information for the manuscript ``Overscreening and Underscreening in Solid-Electrolyte Grain Boundary Space-Charge Layers''.
It comprises the following sections:
\begin{enumerate}
  \item Discussion of Fig.~2 from the main manuscript when replotted on a semi-logarithmic scale.
  \item Discussion of the choice of non-dimensional coupling parameter used in Fig.~3 of the main manuscript.
  \item Further details of the Bayesian analysis of the simulated space-charge profiles and subsequent parameter estimation.
\end{enumerate}
\end{abstract}

\maketitle

\section{Discussion of Fig.~2 from the main manuscript when replotted on a semi-logarithmic scale}

\begin{figure*}[tbh!]
  \centering
  \resizebox{17.8cm}{!}{\includegraphics*{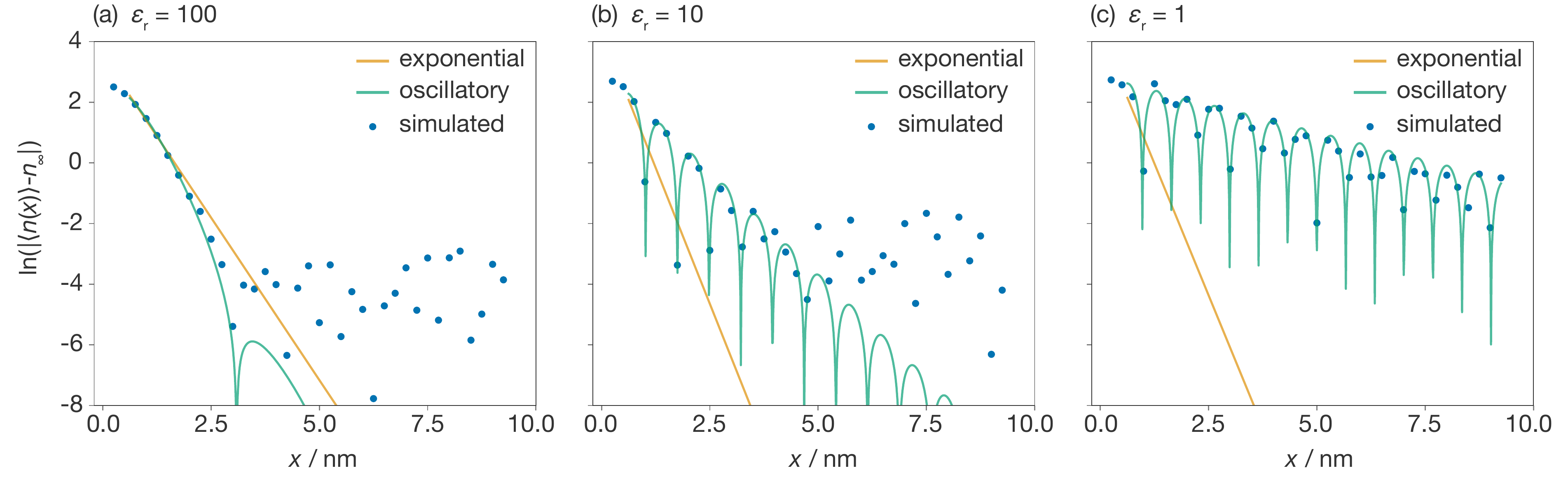}} %
    \caption{\label{fig:overscreening_logarithmic}$\ln\left(\left|\left<n(x)\right>-n_\infty\right|\right)$ for $n_{\infty} = 0.005$ and relative permittivities (a) $\rperm =100$, (b) $\rperm=10$, and (c) $\rperm=1$.
    For each set of simulation data we also plot the maximum likelihood exponential and oscillatory models. 
    \emph{Source:} Raw simulation data and scripts to generate this figure are available under CC BY 4.0 / MIT licences as Ref.~\onlinecite{Dean_dataset}.}
\end{figure*}

In Fig.~2 of the main manuscript we present the one-dimensional time-averaged mobile defect distributions for $n_{\infty} = 0.005$ and relative permittivities (a) $\rperm =100$, (b) $\rperm=10$, and (c) $\rperm=1$, plotted on a linear scale.
For $\rperm = 10$ and $\rperm=1$ (Figs.~2b and 2c) the damped oscillatory decay model gives qualitatively better agreement with the simulation data than the purely exponential decay model.
For $\rperm = 100$ (Fig.~2a) the two models give similar space-charge profiles, which both agree qualitatively with the simulation data.
To better show the differences between these two models, we have replotted the data from Fig.~2 on a semi-logarithmic scale at all three relative permittivities (Fig.~\ref{fig:overscreening_logarithmic}).
The oscillatory behaviour of the simulated space-charge profiles is now more easily visible, even at $\rperm=100$.
This preferential agreement with the oscillatory model, even at this high relative permittivity, is also supported by the calculated Bayes factor for this simulation of $B=196$ (indicating strong support for the more complex damped oscillatory model).
In Figs.~\ref{fig:overscreening_logarithmic}b and Figs.~\ref{fig:overscreening_logarithmic}c the replotted data highlights that the long-ranged decay of these oscillatory space-charge profiles is still exponential (following a straight line on this semi-log plot).

\section{Discussion of the choice of non-dimensional coupling parameter used in Fig.~3 of the main manuscript.}

\begin{figure}[tb]
  \centering
  \resizebox{7.0cm}{!}{\includegraphics*{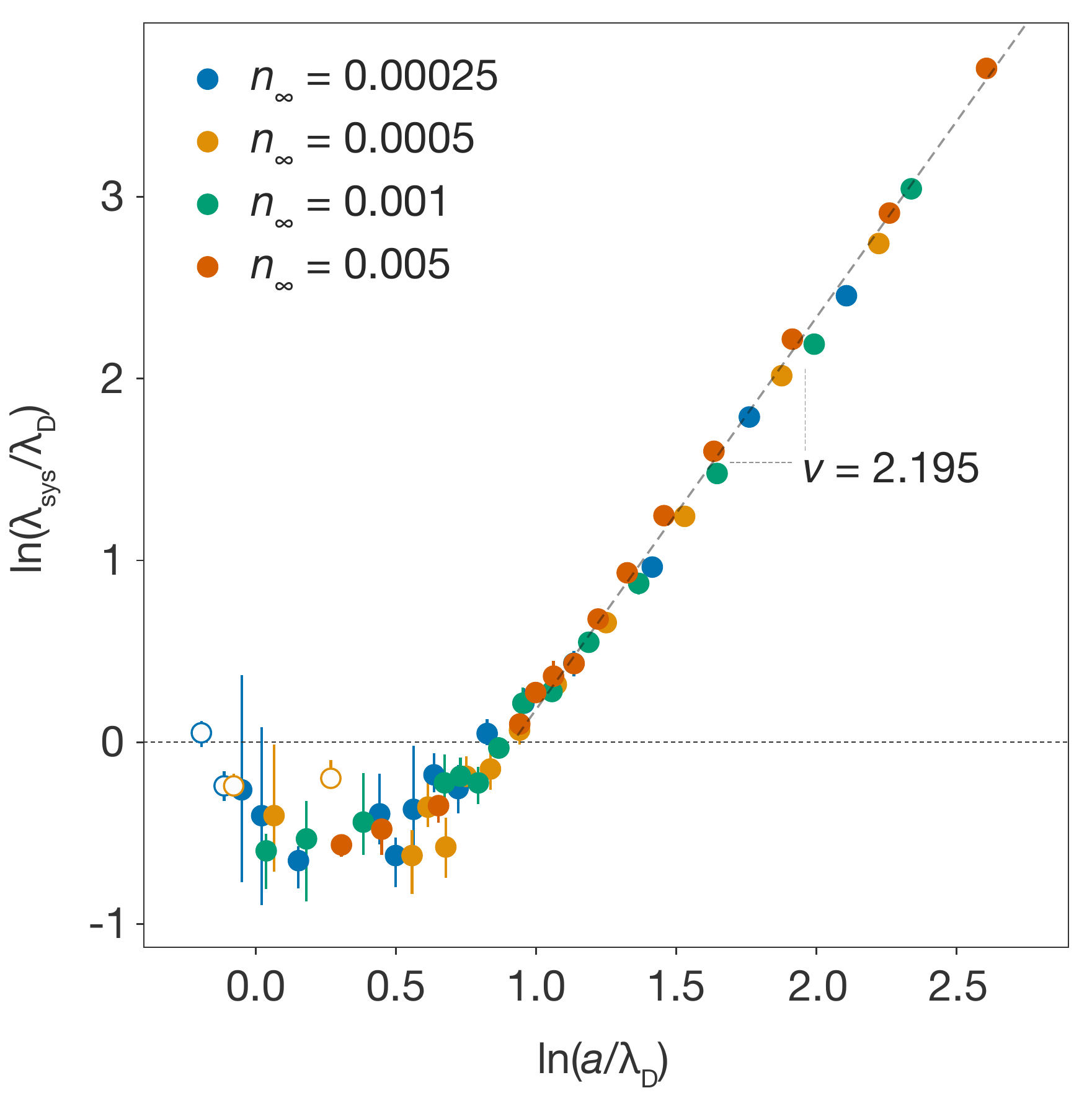}} %
    \caption{\label{fig:a_ws} Log--log plot of scaled simulated decay length $\lambda_\mathrm{sys}/\lambda_\mathrm{D}=1/(\alpha\lambda_\mathrm{D})$ versus $a/\lambda_\mathrm{D}$ for simulations performed at four different defect concentrations.
    The diagonal dashed line shows the maxiumum likelihood estimate for Eqn.~4 in the underscreening regime.
    Open and closed circles correspond to space-charge profiles that best support the pure exponential decay model or the oscillatory decay model respectively, as determined through Bayesian model selection (see below for details).
    Error bars show 95\% confidence intervals for $\lambda_\mathrm{sys}/\lambda_\mathrm{D}$.
    \emph{Source:} Raw simulation data and scripts to generate this figure are available under CC BY 4.0 / MIT licences as Ref.~\cite{Dean_dataset}.}
\end{figure}

In Fig.~3 of the main manuscript we present the variation in simulated screening length, normalised with respect to the corresponding Debye length, as a function of the parameter $\Gamma=\lambda_\mathrm{B}/a$, where $\lambda_\mathrm{B}=e^2/(4 \pi \varepsilon_0 \varepsilon_\mathrm{r} k_\mathrm{B} T)$ is the Bjerrum length, and $a=\left[3/(4\pi n_\infty)\right]^{1/3}$ is a Wigner--Seitz defect-sphere radius \cite{Carley_PhysRev1963,BrushEtAl_JChemPhys1966}.
$\Gamma$ provides a non-dimensionalised measure of the strength of electrostatic coupling between mobile defects \cite{BausAndHansen_PhysRep1980}.
The choice of $\Gamma$ for our analysis is inspired by its historical use in  studies of one-component plasmas \cite{BrushEtAl_JChemPhys1966,BausAndHansen_PhysRep1980}; these include a number of studies of one-component Coulomb lattice-gas models, which correspond to the bulk analogue of the grain boundary model we investigate here.
$\Gamma$ also provides an intuitive sense of how changes in model parameters---defect concentration, $n_\infty$, relative permittivity, $\rperm$, or temperature, $T$---are expected to change the relative strength of electrostatic coupling.
The scaling of $\Gamma$ with respect to $n_\infty$, $\rperm$, and $T$ is given by
\begin{equation}
    \Gamma = \frac{\lambda_\mathrm{B}}{a} \propto \frac{n_\infty^{\frac{1}{3}}}{\rperm T}.
\end{equation}
We therefore obtain the intuitive result that increasing defect concentration, or decreasing either the relative permittivity (increasing charge--charge screening) or the temperature will increase the electrostatic coupling strength.
The form of $\Gamma$ also highlights that we can expect to be in the weakly coupled regime if $a>>\lambda_\mathrm{B}$, i.e.~the typical defect--defect distance is larger than the distance at which defect--defect interactions are comparable to $k_\mathrm{B}T$, and the effect of Coulombic interactions can be treated as a mean-field perturbation \cite{BausAndHansen_PhysRep1980}.
Conversely, we can expect to be in the strongly coupled regime if $a<<\lambda_\mathrm{B}$, i.e.~the typical defect--defect distance is shorter than the distance at which the electrostatic defect--defect interactions are comparable to $k_\mathrm{B}T$.

We note that other choices of dimensionless scaling parameter with equivalent scaling behaviour would give the same phenomenological behaviour when plotted on a double-logarithmic scale.
For example, one analogous scaling parameter that might be considered to be closer in spirit to $d/\lambda_\mathrm{D}$, where $d$ is an ion diameter, used to describe underscreening in ionic liquids and liquid electrolytes, is $a/\lambda_\mathrm{D}$.
This scaling parameter scales with respect to $n_\infty$, $\rperm$, and $T$ as
\begin{equation}
    \frac{a}{\lambda_{D}} \propto \frac{n^{\frac{1}{6}}}{(\rperm T)^\frac{1}{2}}.
\end{equation}
i.e.
\begin{equation}
  \Gamma \propto \left(\frac{a}{\lambda_\mathrm{D}}\right)^2.
\end{equation}
A log--log plot of $\lambda_\mathrm{sys}/\lambda_\mathrm{D}$ as a function of $a/\lambda_\mathrm{D}$  (Fig.~\ref{fig:a_ws}) therefore shows the same universal scaling behaviour as plotting against $\Gamma$.
As expected from the scaling analysis above, the scaling parameter $\nu$ obtained when using $a/\lambda_\mathrm{D}$ is related to the parameter obtained when using $\Gamma$ by a factor of $2$.
The gradient of the linear region in Fig.~\ref{fig:a_ws}, calculated using the same Bayesian analysis procedure used for Fig.~3, is $2.17\pm0.01$ (95\% \textrm{CI}), twice that of the gradient calculated in the main text where $\Gamma$ is used.

\section{Further details of the Bayesian analysis of the simulated space-charge profiles and subsequent parameter estimation}

The statistical modelling of space charge profiles (Fig.~2 and Fig.~\ref{fig:overscreening_logarithmic}) and the scaling law (Fig.~3 and Fig.~\ref{fig:a_ws}) in the main paper was performed using Bayesian analysis. 
For the data presented in Fig.~2 and Fig.~\ref{fig:overscreening_logarithmic}, we have performed maximum likelihood sampling to obtain posterior distributions for the model parameters (the use of a uniform prior probability means that the maximum likelihood parameters and the maximum \emph{a posteriori} values are equivalent). We have then performed nested sampling to estimate the Bayesian evidence $B$ for the more complex damped oscillatory model versus the simpler purely exponential model. To estimate the scaling parameters for the data presented in Fig.~3 and Fig.~\ref{fig:a_ws} we used maximum likelihood sampling. 

{\renewcommand{\arraystretch}{1.5}
\begin{table*}[b]
\caption{Bounds used in Bayesian analysis calculations for each of the four concentrations considered. Curves were fit to distributions of average number of charges in a given plane, not the average fractional site-occupation number $\langle n(x) \rangle$. }
\label{tab: bounds}
\footnotesize
\centering
\begin{tabular}{|l|c|c|c|c|c|c|c|c|c|c|}
\hline
\multirow{2}{*}{} & \multicolumn{2}{c|}{$\alpha$} & \multicolumn{2}{c|}{$A$} & \multicolumn{2}{c|}{$n_{\infty}$} & \multicolumn{2}{c|}{$\xi$} & \multicolumn{2}{c|}{$\theta$} \\ \cline{2-11}
                  &     lower      &    upper      &    lower       &       upper   &   lower        &      upper    &      lower     &     upper     &     lower      &    upper      \\ 
\hline
        $n_{\infty} = 0.00025$          &    $10^{6} $       &    $10^{10} $      &     0      &    10      &      0.9     &    1.9      &      $10^{6} $     &    $10^{10} $      &     0      &   2$\pi$       \\ 
\hline
        $n_{\infty} = 0.0005$          &      $10^{6} $     &      $10^{10} $    &      0     &    25      &     2.3      &      3.3    &      $10^{6} $     &      $10^{10} $    &       0    &    2$\pi$      \\ 
\hline
         $n_{\infty} = 0.001$         &      $10^{6} $     &      $10^{10} $    &     0      &      35    &      5.1     &    6.1      &     $10^{6} $      &    $10^{10} $      &     0      &  2$\pi$        \\ 
\hline 
          $n_{\infty} = 0.005$        &      $10^{6} $     &     $10^{10} $     &     0      &    75      &     26      &     30     &     $10^{6} $      &     $10^{10} $     &    0       &  2$\pi$      \\ 
\hline
\end{tabular}
\end{table*}}

\subsection{Space charge profile modelling}
The Bayesian analysis of the exponential and oscillatory models used uniform prior parameter distributions. 
These bounding values for these distributions are given in Table~\ref{tab: bounds} for each bulk defect concentration. 
The prior distributions used for both models are identical for $\alpha$, $A$, and $n_{\infty}$ at a given concentration. 
Additionally, we have calculated Bayes factors for the comparative evidence supporting the two proposed space-charge models (damped oscillatory versus purely exponential), where the prior probability of both models was \num{0.5}.

Because of the size of our simulations is limited by computational cost, the time-average space-charge profiles, $\left<n(x)\right>$, are not guaranteed to relax to the exact nominal bulk defect concentration, $n_\infty$.
When fitting the exponential and oscillatory model functions to our simulation data, we therefore consider $n_\infty$ as a free fitting parameter.
The layers immediately adjacent to the grain-boundary $xy$ plane do not follow the same asymptotic behaviour as the space-charge profile. We therefore exclude three layers for the $n_\infty = 0.00025$, and $n_\infty=0.0005$ distributions and two layers for the $n_{\infty} = 0.001$, and $n_\infty= 0.005$ distributions from our fit to get a more accurate description of the long-range behaviour.

The use of nested sampling estimates the Bayesian evidence, $\mathcal{Z}$, for a given model with respect to the data. 
This evidence can then be used to calculate the Bayes factor, $B_{01}$, between the null hypothesis model $0$ and the more complex model $1$, 
\begin{equation}
    2\ln{B_{01}} = 2(\ln{\mathcal{Z}_{x}} - \ln{\mathcal{Z}_{y}}).
\end{equation}
Individual Bayes factors were interpreted using the classification given by  Kass and Raftery \cite{KassAndRaftery_JAmerStatAssoc1995}, where $2\ln{B_{01}} > 10$ (defined as ``very strong'' support for the more complex model) was considered sufficient use of the damped oscillatory model for subsquently estimating $\alpha$. 
The evidence values and resulting Bayes factors for each of the space charge profiles are provided in the supporting dataset as Ref.~\onlinecite{Dean_dataset}.

\subsection{Scaling law modelling}
Uniform prior parameter distributions were also used in the maximum likelihood sampling of the scaling law.
The constant of proportionality was bounded by $(-1, +1)$ and $\nu$ was bounded by $(0, +5)$.

\bibliographystyle{ieeetr}
\bibliography{Jacob}